\documentstyle[draft,epsf]{mn}

\title[Regularities in Galaxy Spectra]
       {Global regularities in integrated galaxy spectra}
\author[L. Sodr\'e Jr. \& H. Cuevas]
       {Laerte Sodr\'e Jr.$^{1}$\thanks{e-mail: laerte@astro1.iagusp.usp.br}  
    \&  H\'ector Cuevas$^{1}$\thanks{e-mail: hcuevas@astro1.iagusp.usp.br} \\ 
$^1$Departamento de Astronomia, Instituto Astron\^omico e Geof\'\i sico da 
USP, Av. Miguel Stefano 4200, 04301-904 S\~ao Paulo, Brazil \\
}
\pagerange{\pageref{firstpage}--\pageref{lastpage}}
\def\etal{{et al.~}}
\def\eg{{e.g.~}}
\def\ie{{i.e.}}
\newbox\grsign \setbox\grsign=\hbox{$>$} \newdimen\grdimen \grdimen=\ht\grsign
\newbox\simlessbox \newbox\simgreatbox
\setbox\simgreatbox=\hbox{\raise.5ex\hbox{$>$}\llap
     {\lower.5ex\hbox{$\sim$}}}\ht1=\grdimen\dp1=0pt
\setbox\simlessbox=\hbox{\raise.5ex\hbox{$<$}\llap
     {\lower.5ex\hbox{$\sim$}}}\ht2=\grdimen\dp2=0pt

\def\simless{\mathrel{\copy\simlessbox}}
\newbox\simppropto
\setbox\simppropto=\hbox{\raise.5ex\hbox{$\sim$}\llap
     {\lower.5ex\hbox{$\propto$}}}\ht2=\grdimen\dp2=0pt

\begin{document}

\label{firstpage}

\maketitle

\begin{abstract}
We have investigated some statistical properties of integrated spectra of
galaxies from Kennicutt (1992a) spectrophotometric atlas.
The input for the analysis are galaxy spectra sampled in 1300 bins
between 3750 \AA~ and 6500 \AA. We make use of Principal Component Analysis
(PCA) to analyse the 1300-dimensional space spanned by the spectra.
Their projection onto the plane defined by the first two principal
components, the principal plane, 
shows that normal galaxies are in a quasi-linear 
sequence that we call spectral sequence. We show that the spectral 
sequence is
closely related to the Hubble morphological sequence. 
These results are robust in the sense that the reality
of the spectral sequence does not depend on data normalization.
The existence of this sequence suggests
that a single parameter may describe the spectrum of normal galaxies.
We have investigated this hypothesis with Bruzual \& Charlot (1995) models
of spectral evolution. We show that, for single age models
(15 Gyr), the spectral sequence can be 
parametrized by the characteristic star formation time-scales of 
the different morphological types.
By examining the projection of evolutionary tracks of normal galaxies onto
the principal plane, we verify that the spectral sequence is also an evolutive
sequence, with galaxy spectra evolving from later to earlier spectral types.
Considering the close correspondence between the spectral and 
morphological sequences, this lead us to speculate that galaxies may evolve 
morphologically along the Hubble sequence, from Sm/Im to E.

\end{abstract}


\section{Introduction}

Normal galaxies tend to present great morphological regularities, which allow 
us to classify them along the Hubble sequence (Hubble 1926, Sandage 1961).
The integrated spectra of nearby galaxies also show remarkable regularities.
Their general properties have been recently discussed by Kennicutt (1992a,b),
who has shown that normal galaxies have spectra
that progress smoothly with the morphological type.
Since the integrated spectrum of a galaxy represents a weighted mean in
luminosity of the stellar populations that make it up, these results indicate
that galaxies of same morphological type tend to have similar stellar
populations, what provides the basis for a spectral classification of galaxies
(Humason 1936, Morgan \& Mayall 1957).

We present here a study of Kennicutt's (1992a) spectrophotometric 
atlas, looking for global regularities in the integrated spectra of galaxies. 
This is done with a standard statistical technique, Principal Component 
Analysis (PCA). PCA has previously been applied to the analysis and 
objective classification of 
spectra. Francis \etal (1992) have carried out a study of a large sample of
QSO spectra, developing a quantitative classification scheme based on the 
first three principal components. Sodr\'e \& Cuevas (1994) have analysed a 
set of 24 normal galaxies from Kennicutt's atlas, concluding that most of the 
variance present in their integrated optical spectra is due to their 
morphological differences. They have shown that these spectra may be
parametrized by the morphological type, what allows a 
quantitative galaxy classification based on integrated spectra. 
Connolly \etal (1995) have studied the continuum spectra of the central regions
of ten galaxies, covering a wavelength range from 1200\AA~ to 1$\mu$m, also
finding that ordinary galaxy spectra can be described by a one-parameter 
family. Folkes, Lahav \& Maddox (1996) have used a combination of PCA and 
artificial neural networks for automated classification of galaxies from 
low signal-to-noise spectra.

Here we describe each spectrum by a point in a high-dimensional space where 
each dimension is the flux at each wavelength. 
We then apply PCA to obtain a suitable projection of the spectra onto a plane.
This is done by identifying the orthogonal combinations of the variables with 
maximum variance, the \lq principal components'. Such a projection provides
a synthetic view of the data space, allowing to investigate correlations
between the spectra.

The plan of this paper is as follows. In section 2 we describe our input data,
taken from Kennicutt's atlas, and a data subset containing only
normal galaxies. Some aspects of principal component analysis that are useful 
for the study presented here are reviewed in
section 3. In this section we also discuss data normalization and scaling,
since it is well known that it affects the output of PCA. The results of the
application of PCA to all galaxies, as well as to the subset of normal 
galaxies, are presented in section 4 and discussed in section 5. Finally, 
section 6 summarises our conclusions.

\section{\bf The data}
The 55 integrated spectra investigated here are
from Kennicutt's (1992a) spectrophotometric atlas of galaxies. Most of the
observations cover the wavelength range between 3650 \AA~ and 7100 \AA, with
5 - 8 \AA~ resolution. To avoid incompleteness in the spectral coverage, we
analyse here the rest-frame wavelength interval 3750 - 6500 \AA. This spectral
range was then uniformly re-sampled with 1300 bins, in order to approximately
preserve the original bin width. The spectra in the atlas are
normalized to unity at 5500 \AA, implying that we are not taking into account 
any dependence with galaxy luminosity. We discuss below other normalizations 
and their effect on the results.

We have analysed two sets of data, one containing all galaxies (${\cal AGS}$),
and a subset of ${\cal AGS}$ with normal galaxies only (${\cal NGS}$). 
The sample of normal galaxies contains 23 objects and covers the Hubble 
sequence from E to Im, avoiding objects with any evidence of peculiarity 
(\eg AGNs, starbursts, mergers). The galaxies in this set are:
NGC3379 (E0), NGC4472 (E1/S0), NGC4648 (E3), NGC4889 (E4), NGC3245 (S0),
NGC3941 (SB0/a), NGC4262 (SB0), NGC5866 (S0), NGC1357 (Sa), NGC2775 (Sa),
NGC3368 (Sab), NGC3623 (Sa), NGC1832 (SBb), NGC3147 (Sb), NGC3627 (Sb),
NGC4775 (Sc), NGC5248 (Sbc), NGC6217 (SBbc), NGC2903 (Sc), NGC4631 (Sc),
NGC6181 (Sc), NGC6643 (Sc), and NGC4449 (Sm/Im).
Although small, the ${\cal NGS}$ allows us to explore the connection between 
normal galaxy spectra and morphology.

\section{Principal Component Analysis}
\subsection{General Principles}
Suppose we have a sample of N integrated spectra of galaxies, all covering the
same rest-frame wavelength range. 
Each spectrum is described by a $M$-dimensional vector
${\bf X}$ containing the galaxy flux at $M$ uniformly sampled wavelengths.
Let ${\cal S}$ be the $M$-dimensional space spanned by the `spectral' vectors
 ${\bf X}$.
An integrated spectrum, then, is a point in ${\cal S}$-space, and the
spectra in the sample form a cloud of points in ${\cal S}$.
If the spectral vectors of normal galaxies are indeed correlated with their
morphology, one would expect that these points would be arranged more or less
along a line, mimicking the Hubble morphological sequence. 
It is impossible, of course, to visualize how the data is distributed in
high-dimensional spaces, like the 1300-dimensional spectral 
space discussed in next section. An alternative is to employ some
technique  for dimension reduction by projecting the data in, say, two 
dimensions.

Here we analyse the data with a standard technique, Principal Component 
Analysis (PCA). It is also known as Hotelling  transform,
or discrete Karhunen-Lo\`eve transform, but we call it PCA because this
designation is more often used in Astronomy (\eg Murtagh \& Heck 1987 
and references therein). 

PCA is an orthogonal transformation that 
allows building more compact linear combinations of the data that are optimal
with respect to the mean square error criterion. 
A detailed description of PCA can be found in several books on statistics or
pattern recognition (\eg Kendall 1975, Fukunaga 1990), as well as in the
astronomical literature (\eg Brosche 1973, Whitney 1983, Efstathiou \& Fall 
1984, Lahav \etal 1996). 
Here we restrict ourselves to a short outline of the procedure,
emphasizing only the aspects that are relevant for our analysis.

Consider a set of $N$ objects (galaxies) each with $M$ features (flux 
at $M$ given wavelengths).
Let  $x_{ij}$ be the spectrum of the $i$-th object, \ie, it is the flux
(or a scaled version of it, see below) at the $j-$th wavelength.
It is often more useful to do the analysis with a pre-processed version of 
the data. For example, it is convenient to subtract the sample mean from each
spectrum. This is equivalent to put the origin of the coordinate system at the 
barycentre of the data in the spectral space ${\cal S}$. 
It may be interesting, additionally, to re-scale each variable 
to unit variance. The first case is analogous to PCA of the
covariance matrix, while the second is the same as PCA of the correlation
matrix. 

Let us then assume that the spectral vectors ${\bf X}$ have zero mean. 
The covariance matrix of the data in this case is 
\begin{equation}
C_{jk}=\frac{1}{N-1}\sum_{i=1}^N x_{ij} x_{ik}
\end{equation}
Now, let us consider a new vector, ${\bf Y}$, which is a transformed version 
of ${\bf X}$, given by  
\begin{equation}
{\bf  Y =  A X}
\end{equation}
where ${\bf A}$ is a $M \times M$ matrix whose rows are the eigenvectors of 
the covariance matrix ${\bf C}$.

This transform has several interesting properties. The covariance matrix of
${\bf Y}$ is diagonal, with elements equal to the eigenvalues 
$\lambda_k$ of ${\bf A}$. This means that the transformed vector components
$y_k$ are uncorrelated. Additionally, each eigenvalue $\lambda_k$ is equal to
the variance of the $k$-th element of ${\bf Y}$. Since ${\bf A}$ is a real and
symmetric matrix, 
its inverse is equal to the transpose, ${\bf A^{-1} = A^\prime}$. 
Then, it follows that ${\bf X}$ can be reconstructed from ${\bf Y}$ by using 
the relation
\begin{equation}
{\bf  X = A^\prime Y }
\end{equation}
Suppose, however, that instead of using all the $M$ components of ${\bf Y}$,
we form a new matrix, ${\bf A_K}$, from the $K$ eigenvectors corresponding to 
the $K$ largest eigenvalues. The ${\bf Y}$ vectors will then be 
$K$-dimensional and the reconstruction will no longer be exact. 
Let
\begin{equation}
{\bf {\hat X} = A_K^\prime Y}
\end{equation}
represent the approximation of ${\bf X}$ obtained with the transformation
matrix ${\bf A_K}$. It can be shown that the mean square error between
${\bf X}$ and ${\bf {\hat X}}$ is given by
\begin{equation}
{\cal E} = \sum_{j=1}^M \lambda_j - \sum_{j=1}^K \lambda_j =
\sum_{j=K+1}^M \lambda_j
\end{equation}
This equation indicates that the reconstruction is exact if $K=M$. Also, 
for a given $K$, the error is minimized by selecting the eigenvectors
associated with the $K$ largest eigenvalues. Thus, the PCA transform is
optimal in the mean square error sense.

Note that, contrarily to most implementations of PCA, here we
do not re-normalize each transformed component by its corresponding 
$\lambda_k^{1/2}$, since this gives
unit variance along the new axis, changing the metric of the transformed space.

The PCA transform has an important difference when compared to other orthogonal
transforms: the basis vectors in ordinary orthogonal transforms (\eg  Fourier) 
are fixed, while in PCA they are the eigenvectors of the covariance matrix
and, hence, they are data dependent.

The main body of this paper contains the analysis and discussion of the
projected distribution of the spectra onto the plane $(y_1,y_2)$
defined by the two first components of ${\bf Y}$. We call it the 
{\it principal plane}. 
The first principal component is taken to be along the direction in 
the $M$-dimensional spectral space with the maximum variance.
The second principal component is constrained to lie 
in the subspace perpendicular to the 
first and, within that subspace, it is also taken along the direction with the 
maximum variance. Then, the principal plane is the plane that contains the 
maximum variance in the spectral space. In this sense it is the
most informative plane contained in the data space.

\subsection{Data Scaling}
It is well known that the set of orthonormal coordinates resulting from an 
application of PCA is affected by the scaling of the data, and then
the projection of the spectra onto the plane defined by the first two 
components is different for distinct data normalizations. 
The problem, here, is that the scaling may affect differently
lines and continuum. 

Consider, first, the normalization of the spectra.
Results are different if
the spectra are set to unit at 5500\AA, $f_{5500}=1$, as they appear in 
Kennicutt's atlas, or if they are normalized to have the same mean flux within 
the wavelength range of interest, like 
in the analysis of Francis \etal (1992). 
Connolly \etal (1995) normalize each spectrum to unit
norm, $\sum_\lambda f_\lambda^2 = 1$, and notice that the results are similar
to the latter case above. 
The coefficients of the expansion, in this case, are direction cosines
(e.g. Whitmore 1984), and the spectra are distributed now onto the surface 
of a hypersphere of unit radius in the spectral space.
Here we consider only the normalization of the
spectra to the same 
mean flux, $\sum_\lambda f_\lambda = 1$. We have repeated the analysis with the
other normalizations, verifying that they do not affect our main results.

As discussed in the previous section, the data is usually pre-processed
before the analysis. Flux values at each wavelength can be scaled to zero
mean (which we call covariance method) or to zero mean and
unity variance (correlation method). The variance in the 
lines and continuum are equal in this last case.
In order to control the effects due to scaling, we have applied 
both methods to the analysis of our data sets. 
It is worth pointing out, as we shall see, that our results are also
invariant to the different data scaling, because the linear nature 
of the transform considered here preserves correlations between variables, 
independent of scaling (\eg Francis \etal 1992).

\section{\bf Results}
We have applied PCA to a sample containing 
the spectra of all galaxies in Kennicutt's atlas (set 
${\cal AGS}$), as well as to a subsample comprising only normal galaxies (set 
${\cal NGS}$). The data was scaled before the analysis according to one of 
the procedures discussed above.

We consider first the compression achieved by the two transformed
components $(y_1,y_2)$. The results are summarized in Table 1, which gives
the cumulative fraction of the variance explained by these components for the 
two data sets, and for different scalings. 
We notice that there are not significant differences between the two sets. 
With the covariance method, $\sim$ 93 - 95\% of the total variance 
is described by the two first terms of the expansion. 
For the analysis with the correlation 
matrix, the cumulative variance is lower, 
but at least $\sim 85$\% of the data variance is 
contained in two components now. 

\begin{table}
\caption{Percentage of the variance explained by the first two principal
components}
\label{symbols}
\begin{tabular}{ccc}
method & sample & cumulative variance (\%) \\
                &    &      \\
covariance      & ${\cal NGS}$ & 94.6 \\
                & ${\cal AGS}$ & 93.1 \\
                &    &      \\
correlation     & ${\cal NGS}$ & 86.5 \\
                & ${\cal AGS}$ & 85.0 \\
                &    &      \\
\end{tabular}
\end{table}

\begin{figure}
\centerline{\epsfxsize= 15cm \epsfbox{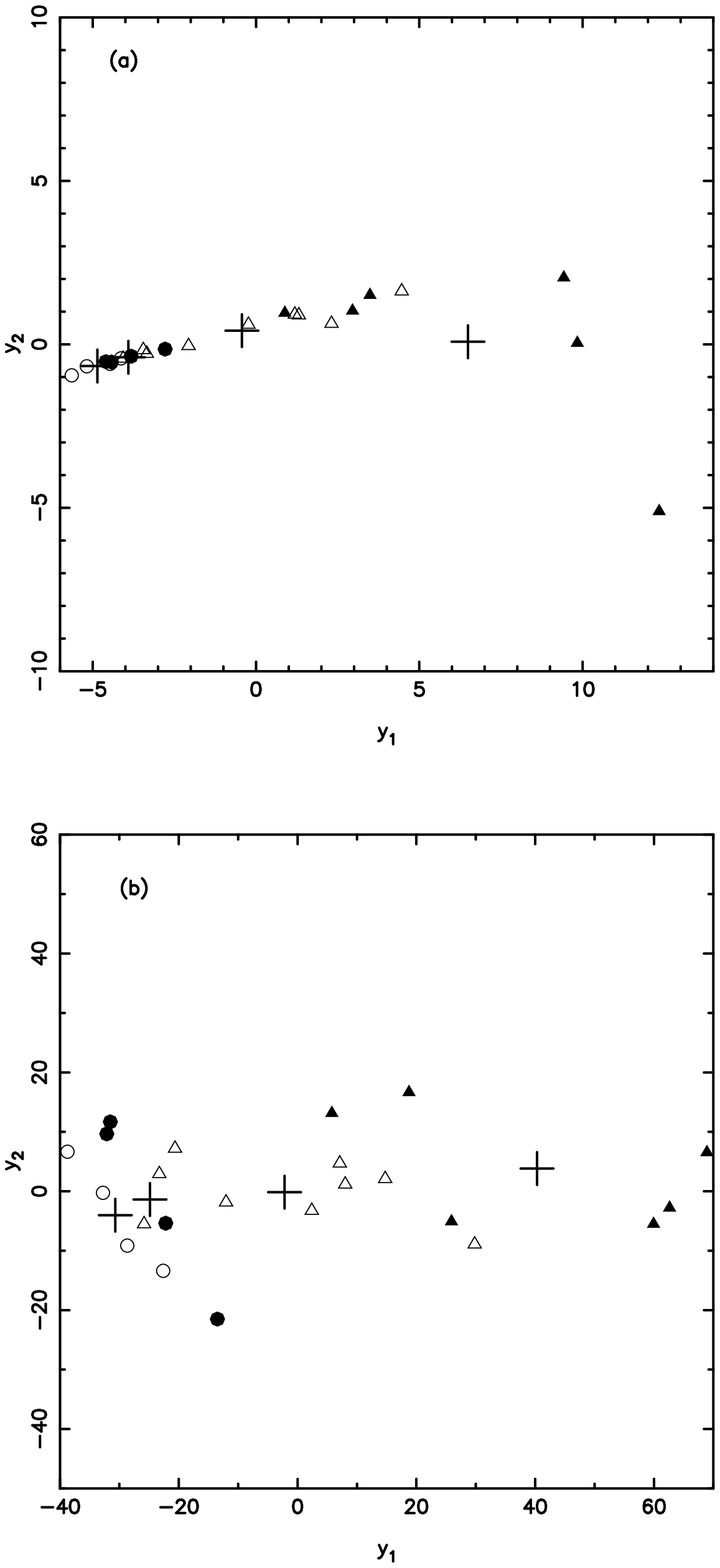}}
\caption{Projection of the spectra of the
${\cal NGS}$ onto the principal plane: (a) covariance method;
(b) correlation method. Different symbols correspond to different morphological
types: E (open circles), S0 (filled circles), Sa-Sbc (open triangles), Sc-Im
(filled triangles).
Also shown in the figure are the projections of
the mean spectra of these 4 morphological groups (crosses): E, S0, Sa-Sbc and 
Sc-Im (from left to right).}
\label{fig1}
\end{figure}
 
\begin{figure}
\centerline{\epsfxsize= 8cm \epsfbox{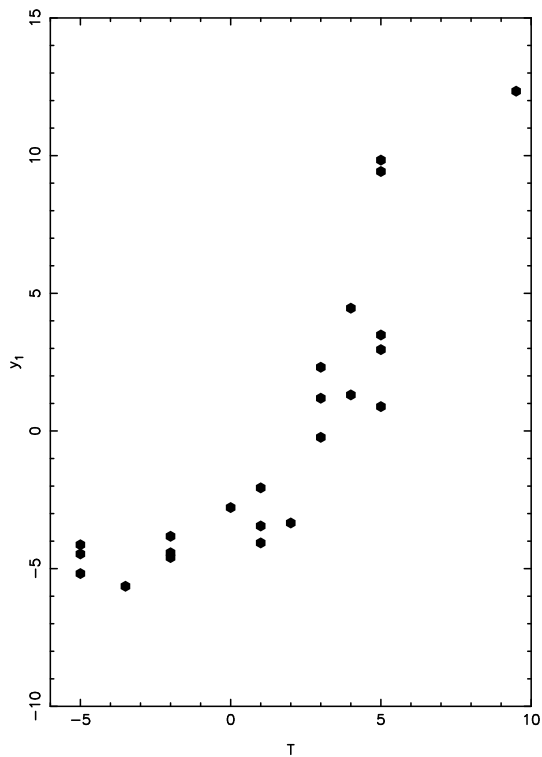}}
\caption{The first principal component, $y_1$, versus
$T$-types for the ${\cal NGS}$ and the covariance method.}
\label{fig2}
\end{figure}

\subsection{\bf Analysis of the normal galaxies set (${\cal NGS}$)}
Figure 1 shows the projection of the spectra of the 
${\cal NGS}$ onto their principal plane. 
The most remarkable aspect in this figure  is that the projected spectra of
almost all galaxies are arranged along a quasi-linear sequence which we shall 
call {\it spectral sequence}. 
There are some outliers of the sequence for the covariance method (figure 1a),
corresponding mainly to late-type galaxies. For instance, the point with 
largest value of $y_1$ in figure 1 corresponds to the Sm/Im galaxy NGC4449.
This is due, at least partially, to the `nebular' nature of the spectrum
of this galaxy (see section 4.3).
Note that the Magellanic irregulars in the atlas are completely
dominated by young stars and HII regions, and their spectra may not be 
representative of this class (Kennicutt 1992a).
The sequence width is relatively
larger for the correlation method (figure 1b), but even 
in this case it is clearly defined (the ratio between the standard deviations
of $y_1$ and $y_2$ is $(\lambda_1 / \lambda_2)^{1/2} \simeq 3.7$).

Most of the variance in the ${\cal NGS}$ seems to be due to the morphological
mix of the sample. We also show in figure 1 the projection of the mean
spectra of 4 morphological groups (E, S0, Sa-Sbc, Sc-Im) onto the principal 
plane as crosses. 
Clearly, these groups are disposed along the spectral sequence keeping
a ranking analogous to the Hubble sequence. It can be verified, however, that
there is significant overlap of spectra of galaxies that are in distinct
morphological groups. Indeed, an inspection of the spectra  shows that
there is a large scatter of spectral properties within each morphological
group, and it is not difficult to find a {\it bona fide} galaxy of a given
morphological type with a spectrum typical of galaxies of another type.
Note that part of this discrepancy should be attributed also 
to the fuzzy nature of the
morphological classification, as discussed by Naim \etal (1995) and Lahav
\etal (1995), who found a dispersion of 1.8 $T$-units in the classification
of a set of 830 APM galaxies by 6 experts. 

In figure 2 we plot the first principal
component $y_1$ versus the Hubble type, measured in the $T$-type 
system of RC3 (de Vaucouleurs \etal 1991), for the covariance method (results
are similar for the correlation method). This plot
presents a clear (non-linear) correlation between $y_1$ and $T$-type. 
In fact, Spearman's rank-order
correlation coefficient between $T$-types and  $y_1$ is
high, 0.93 and 0.90 for the covariance and correlation methods, respectively,
confirming that the correlation 
between these two quantities is significant. The same analysis done with
the second principal component fails to show any significant correlation
between $T$ and $y_2$. We conclude, then, that {\it the spectral sequence
correlates strongly with the Hubble morphological sequence}. This result
is relevant for galaxy classification since it allows to ascribe a type
to a normal galaxy from its spectrum alone (\eg  by its value of $y_1$
or by its position along the spectral sequence).

Another point of interest contained in figure 2 is that the spectral
variation, as measured by $y_1$, is slow from E to Sab, increasing quickly
for later types. This result also is independent of the data scaling,
and reveals that the spectral distance between E ($T$=-5) and Sab ($T$=2)
galaxies is smaller than between Sb ($T$=3) and Sc ($T$=5). 
Kennicutt (1992a) arrived at the same conclusion, arguing that unless one
were able to measure the blue continuum colours or [OII]$\lambda$3727 
emission line to high accuracy, it would be difficult to distinguish the
spectrum of a Sa-Sb from an E-S0 galaxy. Figure 2 also shows that the variance 
of $y_1$ 
increases with $T$, indicating that late-type galaxies present a
richer spectral variety than early-types, at least for the sample studied 
here.

\subsection{\bf Analysis of all galaxies (${\cal AGS}$)}
Figure 3 shows the projection of the ${\cal AGS}$ spectra onto
their principal plane for the covariance method. 
This set includes normal galaxies, AGN, starburst, and interacting or merging
galaxies. Most of the variance,
now, is due to two galaxies only: Mk59 and Mk71. They are Sm galaxies with
spectra dominated by HII-region emission, and here too the nebular nature
of their spectra as well as the faintness 
of their continuum seems to be the cause of their position in the
figure. The point to be stressed, however, is that 
the normal galaxies are again disposed along a sequence.
This is best seen in figure 3b, where we have zoomed out
part of figure 3a.
The projections resulting from the correlation method are shown in figure 4, 
and are qualitatively similar to those displayed in figure 3. The spectral
sequence is clearly visible, what confirms that it is an intrinsic property
of the integrated spectra of normal galaxies.

How are galaxies other than the normal ones distributed in the $(y_1,y_2)$
plane? The answer now depends on the scaling and,
for most types of peculiar galaxies, their location with respect to the
normal galaxies varies from plot to plot in figures 3 and 4. An important 
exception are the starburst nuclei NGC3471, NGC5996 and NGC7714, which 
consistently fall over the Sc-Im region of the spectral sequence. 

\begin{figure}
\centerline{\epsfxsize= 15cm \epsfbox{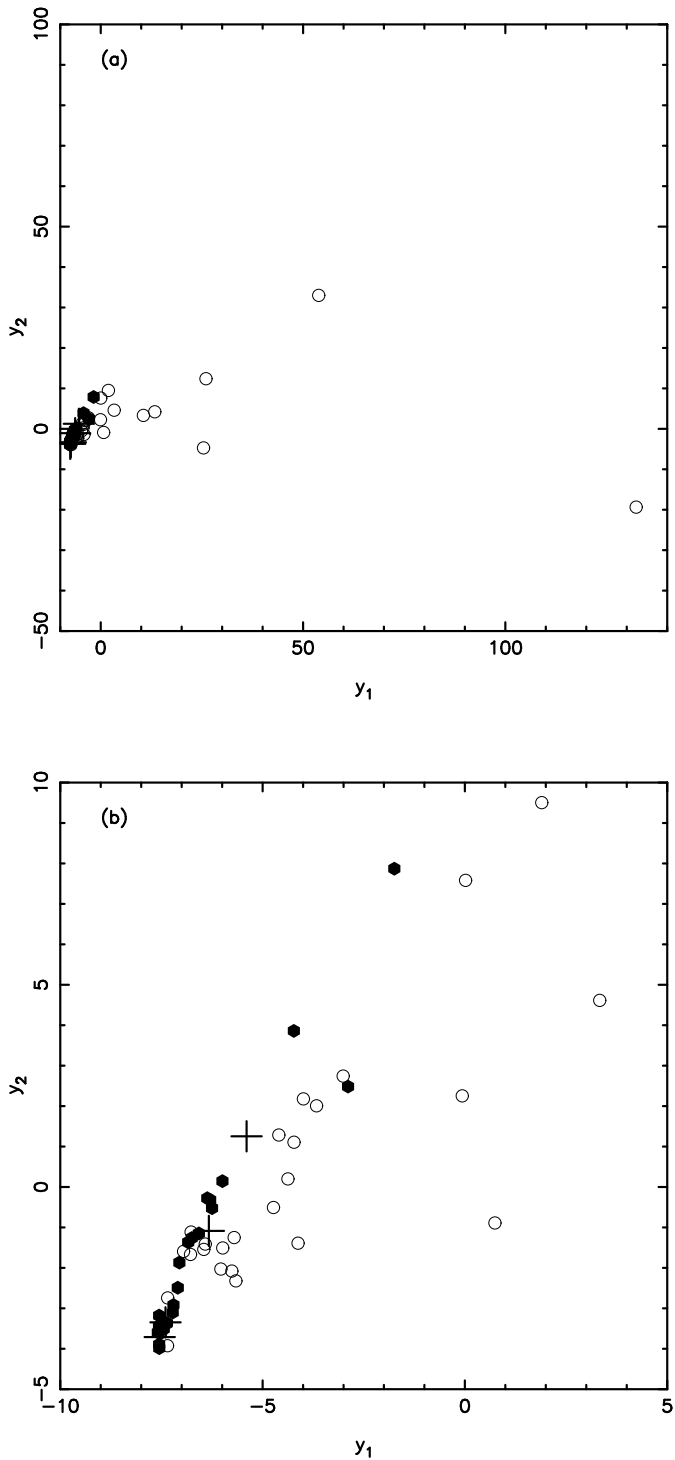}}
\caption{Projection of the spectra of
${\cal AGS}$ onto the principal plane for the covariance method.
The projected spectra of normal galaxies are represented here as filled
hexagons, while those of peculiar galaxies are shown as open circles. The
mean spectra are represented by crosses (see caption of figure 1).
Figure 3b is similar to 3a, but zoomed out to enhance the spectral
sequence.}
\label{fig3}
\end{figure}
 
\begin{figure}
\centerline{\epsfxsize= 10cm \epsfbox{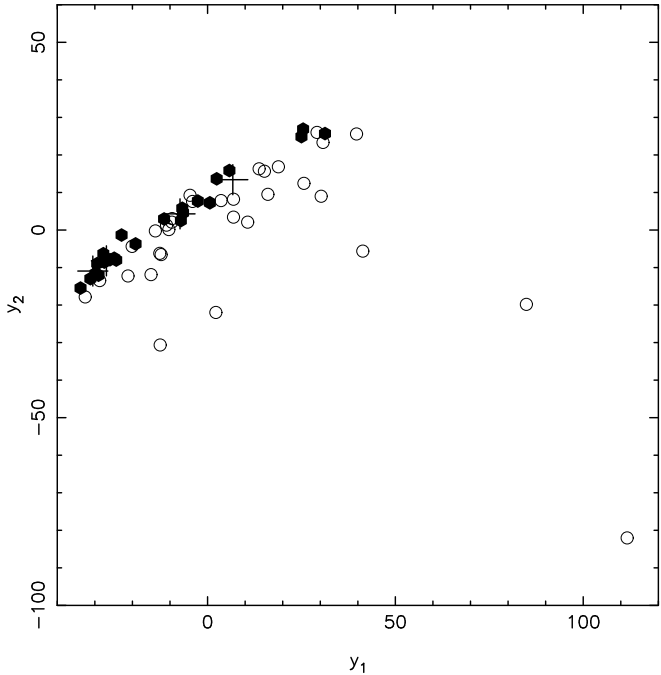}}
\caption{Same as figure 3a, but for the correlation method.}
\vskip 0.5cm
\label{fig4}
\end{figure}

\subsection{\bf The nature of the principal components}
PCA is a linear transform and then each principal component 
is a linear combination of the flux at all wavelengths. 
We plot in figure 5 the weights associated with each wavelength for the two
first principal components for galaxies in the ${\cal NGS}$. This figure is
useful because it provides an insight on what determines $y_1$ and
$y_2$ for each type of data scaling. Results for the ${\cal AGS}$ are similar.

Figure 5a shows the weights corresponding to $y_1$ for the covariance method.
They are the components of the first eigenvector of the covariance matrix.
The overall weight distribution is regular, being positive in the blue and
decreasing to negative values for longer wavelengths. Blue galaxies, then,
tend to have larger values of $y_1$ than the red ones. But
large weights are associated with wavelengths of emission lines, mainly
[OIII]$\lambda \lambda$4959,5007 and H$\beta$.
This might lead one to suppose that $y_1$ is actually measuring the strength of
these lines. To clarify this point, we repeated the analysis excluding small
regions around the Balmer lines and the [OIII] nebular lines.
The projection of the galaxies onto their principal plane looks like
that shown in figure 1b, with the two previous late-type outliers of figure 1a
now disposed along the sequence. Note that most galaxies in the ${\cal NGS}$ 
present negligible or small emission at 
the wavelengths of the [OIII] lines and H$\beta$, and then
the information carried by these wavelengths is proportional to the continuum 
flux at $\sim5000$\AA. Galaxies with strong nebular spectrum will have larger 
values of $y_1$ because they tend to be bluer than the others and, mainly,
because of the large and positive contribution of their line flux.
We conclude that the information associated with the spectral sequence comes
essentially from the continuum, with a significant contribution of emission
lines, when present. This is confirmed by figure 5b, where it is
shown the weights corresponding to $y_2$. These weights are almost zero,
except at the wavelengths close to the emission lines, where they present an
oscillation. These oscillations act like a line detector as it computes a linear
combination of differences between the flux at the line and the nearby 
continuum (adjacent continuum - line, actually):
$y_2$ is near zero for spectra without emission lines, and
negative for those where the lines are prominent. This explains the position
of the  outliers in figure 1a. 

The covariance method preserves the differences between the continuum and the
lines and then our result reveals that a large fraction of the variance of 
the sample is coming from wavelengths associated with H$\beta$ and 
the [OIII] nebular lines. 
In the correlation method, lines and continuum are put to the same variance,
and the resulting weights are quite different, as shown in figures 5c and 5d.
The continuum now provides a contribution to  $y_1$ more significant than the
emission lines. The distribution of the weights shows that $y_1$
is measuring the flux difference between the blue and red parts of a spectrum,
with a non negligible contribution of some lines. The weights associated with
$y_2$ in the correlation method are shown in figure 5d. Their behaviour is
quite different from that shown in figure 5b for the covariance method. Now
$y_2$ is essentially probing the wavelength interval between 4300\AA~ and
5200\AA. Then, to first order, $y_1$ is measuring a colour and 
$y_2$ the relative flux at $\sim$4750\AA.

The discussion in this section reveals that the detailed interpretation of 
the principal components are quite dependent of the data set. 
At the same time, it indicates that is the continuum that gives the major
contribution to the first component, but nebular emission may also be
important.

\begin{figure*}
\begin{minipage}{17.5cm}
\protect\centerline{\epsfxsize= 11cm \epsfbox{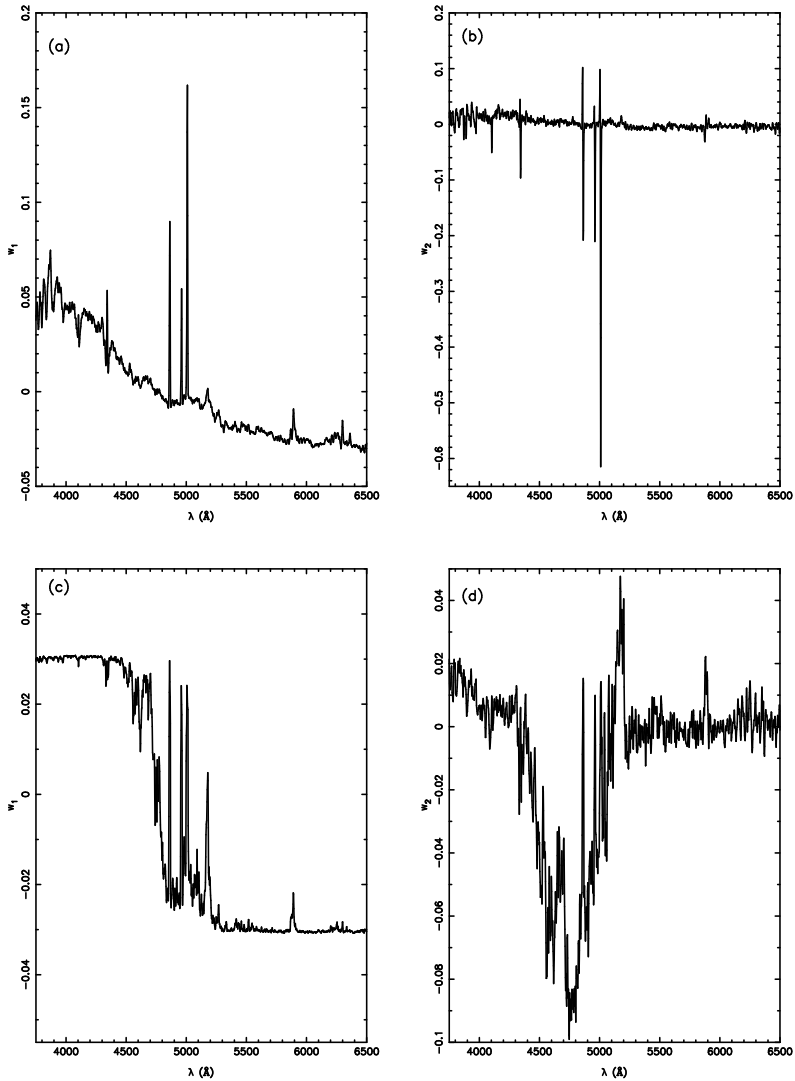}}
\caption{Weights (components of eigenvectors) associated to the
principal components $y_1$ and $y_2$ for the ${\cal NGS}$: 
(a) $y_1$, covariance method; 
(b) $y_2$, covariance method; 
(c) $y_1$, correlation method;
(d) $y_2$, correlation method.}
\label{fig5}
\end{minipage}
\end{figure*}

\subsection{\bf Effect of the errors}
The overall accuracy of the spectrophotometry of the galaxies in the
atlas is $\sim$5-10\%. To investigate the effect of these errors we have
adopted the following procedure. For two normal galaxies, NGC3379 (E0) and
NGC4631 (Sc), we produced 100 simulated spectra by adding Gaussian noise 
with standard deviation equal to 0.1 times the flux at each wavelength to the 
two observed spectra. For each data set and type of scaling,
the simulated spectra were pre-processed and projected onto the 
principal plane.
We have verified that the errors have more effect on the late-type 
galaxy spectrum than on the elliptical spectrum, due to the
emission lines present in the spectrum of the former object. 
Their effect on the results presented here are, however, negligible.

\section{\bf Discussion}
\subsection{\bf The spectral sequence}
The results of the previous section allow us to conclude that the 
spectral sequence is real in the sense that the spectra of normal galaxies 
form a sequence in the spectral space. 
This result is robust because, as we have
seen, it is independent of data normalization and scaling. 
Note that figures 1a and 3 represent projections of the same
data space (for the covariance method) onto different planes: 
while in figure 1a the plane maximizes the variance of the ${\cal NGS}$,
the principal plane in figure 3 maximizes the variance of the ${\cal AGS}$.
The same is true for figures 1b and 4 for the correlation method.

It is well known that several parameters related to spectra, like colours
or spectral indices, correlate well with the Hubble sequence (\eg Roberts
\& Haynes 1994). 
We had already noticed (Sodr\'e \& Cuevas 1994) 
that in a data space defined by spectral parameters,
like the amplitude of the 4000\AA~ break and the strength of the G band or the
 Mg2 index, the morphological types are distributed along 
a one-dimensional sequence. Our results here indicate that these 
empirical correlations are due to the global regularities present in galaxy 
spectra, illustrated here by the spectral sequence.
It provides quantitative support for a one-dimensional
description of the general properties of normal galaxies, like the Hubble
morphological sequence. It also indicates that there is not a dividing line 
in spectral properties between
elliptical and spiral galaxies, and that lenticulars have spectra intermediate
between those two morphological groups (c.f. figures 2-4). 

The existence of the spectral sequence and its relation with the morphological
sequence has been recognized  by Connolly et al. (1995). They applied a
procedure similar to ours to a sample of 10 composite spectra of the central
regions of normal and starburst galaxies, each covering a wavelength range of
1200\AA~ to 1$\mu$m. Interestingly, they found that their 10 spectra
were disposed along a spectral sequence, in disagreement with our results, 
where the spectral sequence is defined only by the normal galaxies. This is 
due to two effects. First, they have removed the strongest nebular emission 
lines from the starburst spectra, decreasing the variance associated with the
lines (c.f. section 4.3). Second, a visual examination of their spectra
indicates that most of the variance, now, is coming from the UV continuum
($\lambda \simless 3000$\AA), a region not included in our analysis. 
This discussion confirms how dependent of the input data the results of PCA 
are! Another point that is worth mentioning is that Connolly \etal 
(1995) found a curved spectral sequence, while in our study it is more or less
linear. This is a consequence of their normalization by the scalar product,
that projects the spectra onto the surface of a hypersphere in the spectral 
space.

Taken together, these results increase the robustness of the spectral 
sequence of normal galaxies, as it appears in all these analysis of galaxy spectra,
irrespective of the wavelength interval, spectral resolution,
or the detailed form of the input.
Consequently, we expect that a study of a sample that includes, say, 
H$\alpha$, will lead to the same general conclusions although the projection
of the galaxies onto the new principal plane probably will be  different.

The principal plane is convenient for classification of normal galaxies,
because we can ascribe an objective spectral type to a normal galaxy from its
position along the sequence. The spectra of
peculiar galaxies, however, occupy positions in the principal plane that 
depend whether one uses the covariance or the correlation method, since the
relative r\^ole of continuum and lines is different for each of the two
scaling  methods. Hence, the
principal plane is not adequate for classification of non-normal galaxies,
\ie, one needs more than just two dimensions to describe the general manifold
of integrated spectra of galaxies. 

An obvious
advantage of spectral classification over morphological classification is
that the former is more adequate to a quantitative approach than the latter.
For instance, Folkes, Lahav \& Maddox (1996) developed a method for galaxy
classification from low signal-to-noise spectra typical of reshift surveys.
They have simulated spectra of normal galaxies with the parameters of the
2dF Galaxy Redshift Survey. Using  a combination of PCA and artificial
neural networks, they show that it is necessary
typically 8 principal components to describe normal noisy spectra and 
be able to classify them. Their results indicate that $\sim95$\% of the
spectra are correctly classified in 5 morphological groups at $b_J=19.7$,
the limiting magnitude of the survey. Since they also used Kennicutt's data,
their PCA results are very similar to ours, even noting that their sample
of normal galaxies contains 3 galaxies more than ours. 

\subsection{\bf Model spectra}
The principal plane provides us with a useful tool to analyse
spectra. 
We illustrate this point with the following exercise. 
The very existence of the spectral sequence, and its correlation with the 
Hubble sequence, indicates that one single parameter
may be responsible for the integrated spectra of normal galaxies- and of
the morphological sequence. For 
instance, Gallagher, Hunter \& Tutukov (1984), 
Sandage (1986) and Ferrini \& Galli (1988),
suggest that several properties of the Hubble sequence can be explained by
variations in the star formation rate of galaxies. 

We have investigated this hypothesis with Bruzual \& Charlot (1995, hereafter
B\&C) revised models of spectral evolution 
of galaxies (see also Bruzual \& Charlot 1993). We have only considered models
with a single parameter, the characteristic star formation 
time-scale of a galaxy, $\tau$, where the SFR decreases with time as
$\exp (-t/\tau)$. We assume a Salpeter IMF,
with lower and upper mass limits equal to 0.1 and 125 M$_\odot$,
respectively, and neglect gas recycling.
Each model results in a spectrum sampled at 238
channels within the wavelength range of interest here. For consistency, we
have computed the principal components of our two data sets using the same 
238 channels of the theoretical spectra, instead of our 1300 previous 
wavelengths. The cumulative variance explained by the
principal plane is not too affected by this reduction of spectral resolution.
For the ${\cal AGS}$, it is now 93.7\% and 86.0\% 
for the covariance and correlation methods, respectively.

\begin{figure*}
\begin{minipage}{17.5cm}
\protect\centerline{\epsfxsize= 15cm \epsfbox{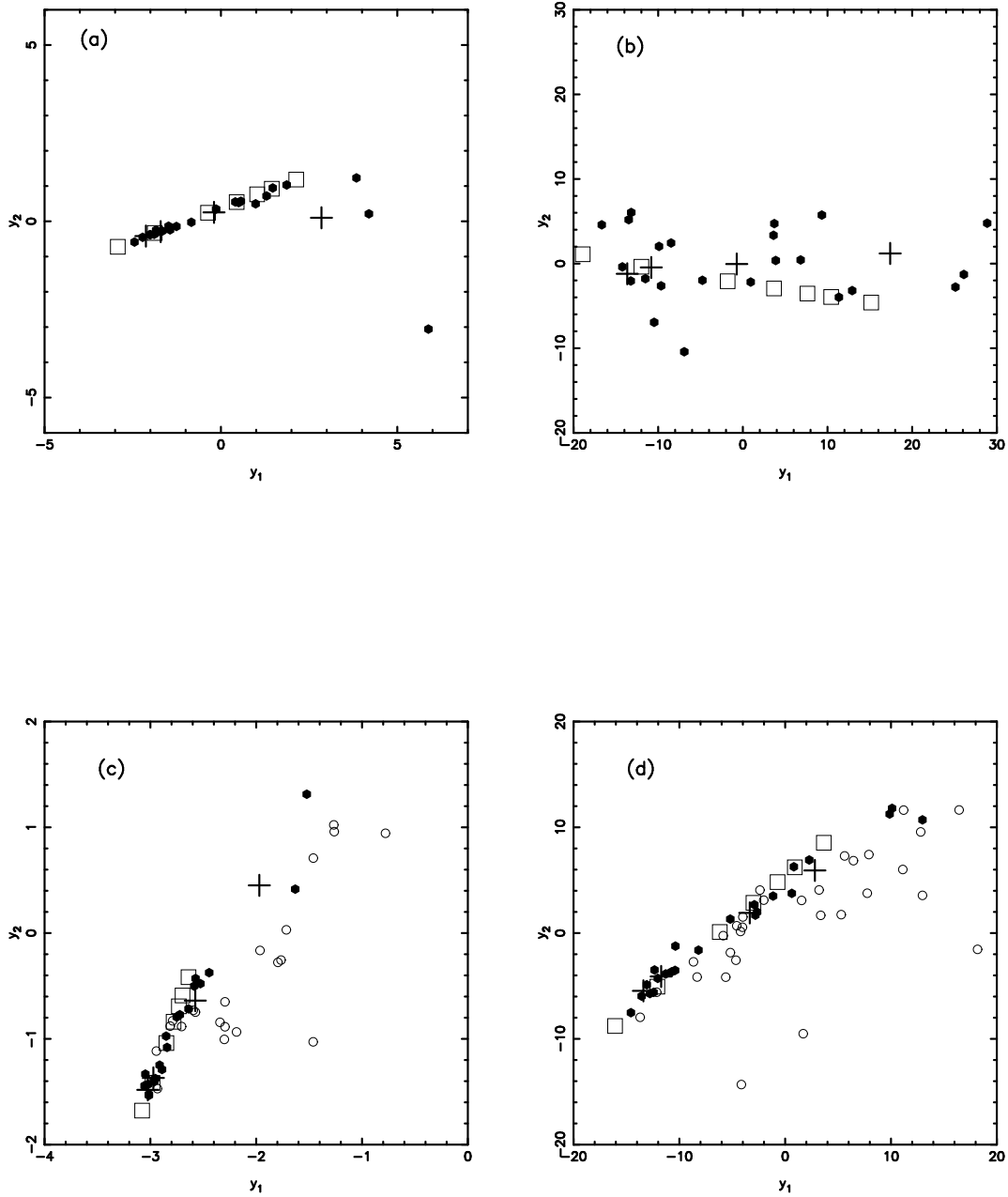}}
\caption{Projection of B\&C model spectra for several values of
the mean star
formation time scale $\tau$ and age equal to 15 Gyr onto the principal plane:
(a) ${\cal NGS}$,
covariance method; (b) ${\cal NGS}$, correlation method; (c) ${\cal AGS}$,
covariance method; (d) ${\cal AGS}$, correlation method. Model spectra
are represented by open squares. They correspond, from left to right, to
$\tau =$0, 3, 5, 7, 10, 15 and $\infty$ Gyr (see text for details).
Other
symbols have the same meaning as in figure 3. In figures 6c and 6d,
only the part of the
principal plane occupied by the spectral sequence is shown.}
\label{fig6}
\end{minipage}
\end{figure*}

Let us first consider single age models (15 Gyr).
In figure 6 we show, for several values of $\tau$, the projection of the
model spectra onto the principal plane defined by the galaxies in each of our
two data sets. The model spectra overlap the spectral sequence for the two
methods, although they tend to fall below the mean types for the correlation
method.  For each data set and normalization,
the spectrum corresponding to a single instantaneous burst 
($\tau=0$, \ie, B\&C model bc95-ssp-sp-1) falls just at the left extreme
of the sequence.
On the opposite extreme, the spectrum of a composite stellar 
population with constant star formation rate 
($\tau = \infty$, \ie, model bc95-cons-sp-1), a simple model
for irregular galaxies,
is near the mean locus of Sc galaxies in the principal plane. Other
values of $\tau$ are distributed along the sequence, between these two
extremes. Figure 6 indicates that the parameter $\tau$ provides a good 
parametrization of the spectral sequence, from elliptical to irregular 
galaxies (although this seems not to be quite true for the correlation
method applied to the ${\cal NGS}$, even in this case the results are
satisfactory if one takes into account the simplicity of the models considered
here). This result is in agreement with the conclusions of 
Kennicutt, Tamblyn \& Congdon (1994),
who have shown that the photometric properties of spiral galaxies
along the Hubble sequence are 
predominantly due to changes in the star formation
histories of disks, and only secondarily to changes in the bulge-to-disk ratio,
which also varies systematically along the morphological sequence.

We plot in figure 7 evolutive tracks for the $\tau=0$ and $\tau=\infty$ 
models, with galaxy ages running from 0 to 20 Gyr. 
The evolutive track of an instantaneous burst, for the covariance method,
overlaps the spectral sequence, that is, its spectrum evolves from late to 
early-types along the sequence. 
The same is partially true for the constant star formation rate 
model, but in this case the oldest spectrum falls near the centroid of normal
Sc galaxies.  Hence, in the principal plane obtained with the
covariance method, the
spectral sequence not only characterizes the locus of normal galaxies but
is almost coincident with their evolutionary tracks.
Figure 7 also shows that, for the correlation method, these  tracks 
present more structure, but the same evolutive trend holds.
It is worth pointing out that these tracks correspond to very simple models,
and `cosmic variance', \eg differences in the epoch of galaxy formation, 
IMF, SFR, metallicity, dust content, etc., may
well lead to models able to cover all the spectral sequence.
Note that these tracks do not explain why some galaxies are out of the spectral
sequence. Although most of them probably 
have a more complex star formation history
than the simple exponential decreasing star formation
rate considered here, the main reason
is that B\&C spectra do not include nebular emission produced by HII regions,
and hence they are not adequate to describe galaxies with spectra 
dominated by ongoing star formation, like Mk59 or Mk71.

\subsection{On the nature of the Hubble sequence}
Our results lead us to conclude that the spectral sequence
is also an evolutive sequence, with galaxy spectra evolving from 
Magellanic irregulars to that of ellipticals. 
Consequently, we expect that the fraction of galaxies with late-type spectra
should increase with redshift. 
It is well known that the fraction of blue objects increases with the redshift,
as evidenced originally by the Butcher-Oemler (1978) effect in clusters 
(see Rakos \& Schombert 1995 for a recent study) or by the dramatic
excess (with respect to no-evolution models) in galaxy counts in faint blue 
magnitudes (see Ellis 1990 for a review). Additionally, recent observations 
with the HST have been able to show that these galaxies are mainly late-type 
spirals and irregulars (Glazebrook \etal 1995; Driver, Windhorst \& Griffiths
1995).

We have seen that there is a close correspondence
between the spectral and morphological sequences today, and
that the spectral sequence is also an evolutive sequence.
If we do not live in a special epoch in the history of the Universe, we may
suppose that the spectral and morphological sequences have been
always related. Then the Hubble sequence itself would be an evolutionary
 sequence, 
with galaxies evolving from Im to E! This is a speculation, of course, since
our results refer to spectra and not to morphologies, and then the appearance
of an E galaxy with age of 1 Gyr may be quite different of what is today an
$\sim$Sc galaxy. But is worth noting that some recent works present evidence
that  the morphology of a normal galaxy may evolve from 
late to early types. For instance, Pfenniger, Combes \& Martinet (1994) and
 Pfenniger, Martinet \& Combes (1996) argue that
this evolution may be driven by internal and external factors due to the
likely coupling between dynamics and star formation. The key point
is that dynamical process that actuate during
a galaxy life- like formation and destruction of bars, mergers, 
close encounters, gas compression and/or stripping, etc.-
tend to  favour an increase of the 
spheroidal component at the expense of the
disk, leading to a univocal sense of 
morphological evolution, from Sm to Sa.
This sense of evolution also explains the morphological content of galaxy
clusters, where most of galaxies are E or S0. The simplest
explanation assumes that there is an infalling population of late-type
galaxies (Sodr\'e et al. 1989, Kauffmann 1995) that are transformed in 
E and S0 (as well as in dwarf ellipticals) in the hostile environment
of the clusters. Moore et al.
(1996) have recently proposed that frequent encounters at high speed among
the galaxies in clusters (``galaxy harassment") may be the driver of
morphological transformations in these environments. This process may explain
the Butcher-Oemler effect and the form of the blue objects observed in
high-redshift clusters by the HST.
Hierarchical models also indicate that galaxy morphology may well change 
as consequence of mergers and interactions (Baugh, Cole \& Frenk 1996),
although these models do not have yet enough resolution to address the question
of evolution within spiral types.  
This scenario also
 suggests that the first galaxies should look more like faint gas
rich irregular objects- in agreement with the inhomogeneous galaxy formation
models of Baron \& White (1987)-
 than the presumed bright precursors of today's
elliptical galaxies and bulges of spirals in the scenario devised by
Eggen, Lynden-Bell \& Sandage (1962).

Note that our results are consistent with 
this `nurture' scenario, but they do not exclude by any means
the `nature' framework, where the morphology of a galaxy is already 
imprinted in the initial conditions at the moment of birth, 
with the galaxy spectrum evolving passively with time,
without any significant morphological evolution.
We hope that HST observations may be able to constrain the amount of 
morphological evolution of high redshift galaxies and help solving the long
standing problem of the origin of the Hubble sequence.

\begin{figure*}
\begin{minipage}{17.5cm}
\protect\centerline{\epsfxsize= 15cm \epsfbox{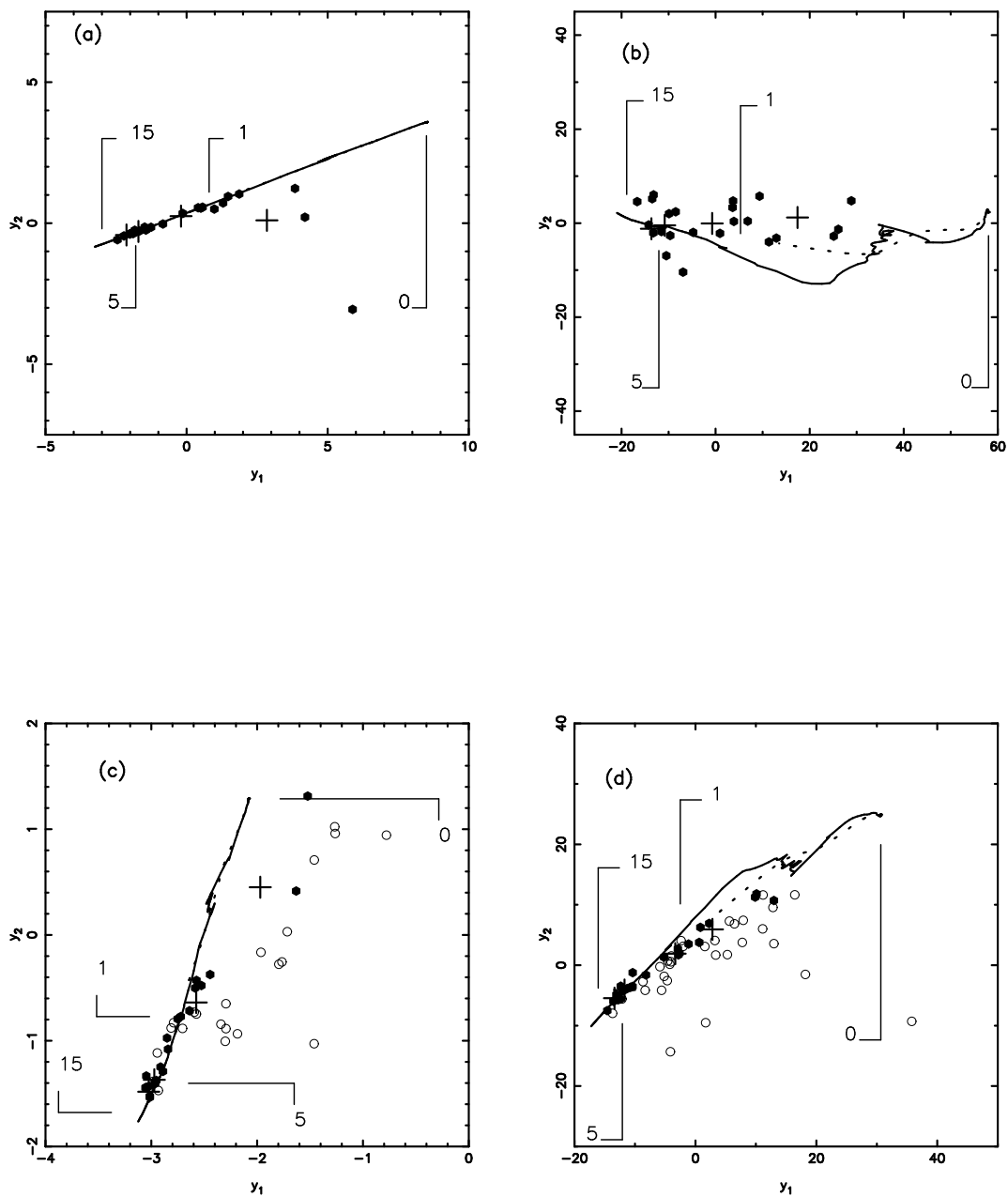}}
\caption{Projection of B\&C evolutive tracks for $\tau=0$
(continuous
line) and $\tau=\infty$ (dotted line) onto the principal plane:
(a) ${\cal NGS}$,
covariance method; (b) ${\cal NGS}$, correlation method; (c) ${\cal AGS}$,
covariance method; (d) ${\cal AGS}$, correlation method.
Symbols have the same meaning as in figure 3. The age (in Gyr) is indicated
next to the tracks. Note that the track for $\tau=0$ completely overlaps
the track for $\tau=\infty$ in the covariance method (7a and 7c).}
 
\label{fig7}
\end{minipage}
\end{figure*}

\section{\bf Conclusions}
We have applied PCA to the study of integrated spectra of galaxies.
We have found useful to analyse the projected distribution of galaxy spectra
onto the principal plane, defined as the plane in the
data space that contains most of the data variance. 
We have shown that normal galaxies delineate a spectral sequence, which is
related to the Hubble morphological sequence, \ie, the spectrum of normal
galaxies changes smoothly from E to S0 to Sa, up to Sc and Im, without a gap 
between ellipticals and spirals. 
The spectral sequence in the principal plane is somewhat analogous to the 
main sequence in the HR diagram. 
The existence of the spectral 
sequence is independent of the data scaling and provides quantitative
support for the Hubble morphological sequence.

The fact that normal galaxies form a one-dimensional spectral sequence 
may indicate that a single parameter controls their integrated spectra 
(and the Hubble sequence). With Bruzual \& Charlot models we have shown 
that the characteristic star formation time scale of galaxies provides a good
parametrization of the spectral sequence. Additionally, the evolutionary
tracks of normal galaxies, when projected onto the 
principal plane defined by the spectra of presently normal galaxies,
overlap completely (for E galaxies) or partially
(for late-type galaxies) the spectral sequence.
This means that galaxy spectra evolve along the sequence,
at least for the simple evolutionary models discussed here.
Taking into account the close correspondence between the spectral and 
morphological sequences, our results are consistent with the hypothesis that
galaxies may also evolve morphologically along the 
Hubble sequence, from Sm/Im to E.

It is worth noting that
PCA is dependent of the data set, since the analysis is done over the
covariance or correlation matrix of the data. As we have shown, however,
our main results are robust in the sense that they are, in a large extent, 
independent of data scaling or spectral resolution. 
It will be interesting to repeat this analysis when richer samples,
including high redshift galaxies, become publically available. 

\section*{ACKNOWLEDGEMENTS}

We are pleased to thank G. Bruzual and S. Charlot for making their galaxy
isochrone synthesis spectral evolution library (GISSEL95) available to us,
Ofer Lahav and Avi Naim for many discussions on PCA, and an anonymous referee
for comments that allowed us to improve the paper.
We also thank Eduardo Telles and Claudia Mendes de Oliveira for their help
on the revision of this paper.
The Kennicutt atlas was obtained from the Astronomical Data Center (ADC).
This work benefited from the financial support pro\-vi\-ded by the Brazilian
agencies FAPESP, CNPq, and CAPES.

\end{document}